\documentclass[aps,prl,twocolumn,groupedaddress,showpacs]{revtex4}

\usepackage{graphicx}
\usepackage{dcolumn}
\usepackage{bm}
\usepackage{amssymb}
\begin{document}

\title{Pure Gas of Optically Trapped Molecules Created from Fermionic Atoms}

\author{S. Jochim}
\author{M. Bartenstein}
\author{A. Altmeyer}
\author{G. Hendl}
\author{C. Chin}
\author{J. Hecker Denschlag}
\author{R. Grimm}

\affiliation{Institut f\"ur Experimentalphysik, Universit\"at
Innsbruck, Technikerstra\ss{}e 25, 6020 Innsbruck, Austria}



\date{\today}

\begin{abstract}
We report on the production of a pure sample of up to
$3\times10^5$ optically trapped molecules from a Fermi gas of
$^6$Li atoms. The dimers are formed by three-body recombination
near a Feshbach resonance. For purification a Stern-Gerlach
selection technique is used that efficiently removes all trapped
atoms from the atom-molecule mixture. The behavior of the purified
molecular sample shows a striking dependence on the applied
magnetic field. For very weakly bound molecules near the Feshbach
resonance, the gas exhibits a remarkable stability with respect to
collisional decay.
\end{abstract}


\maketitle

The formation of composite bosons by pairing of fermions is the
key to many intriguing phenomena in physics, with superfluidity
and superconductivity being prominent examples. In ultracold
atomic gases pairs of fermionic atoms can be combined to form
bosonic molecules \cite{Regal2003a,Cubizolles2003} or possibly
Cooper pairs \cite{Stoof1996a}. The pairing changes the properties
of the gas, highlighted by the prospect of a molecular
Bose-Einstein condensate or a Cooper-paired superfluid. The
interatomic interactions play a crucial role for the nature of the
pairing process. The ability to control the interaction via
magnetically tuned Feshbach resonances
\cite{Feshbach1962a,Inouye1998a,Loftus2002a} opens up exciting
possibilities for experiments on ultracold fermionic gases, e.g.\
exploring superfluidity in different pairing regimes
\cite{Holland2001a,Timmermans2001a,Chiofalo2002,Ohashi2002}.

The formation of molecules near Feshbach resonances in ultracold
gases has been reported for bosons
\cite{Donley2002a,Chin2003a,Herbig2003,Duerr2003} and fermions
\cite{Regal2003a,Cubizolles2003}. In the experiments
\cite{Donley2002a,Chin2003a,Regal2003a,Cubizolles2003}, the
molecules coexist with the atoms in a strongly interacting
mixture. A generic feature of a Feshbach resonance is the
existence of a bound molecular state with a magnetic moment that
differs from that of the unbound atom pair. The binding energy
thus depends on the magnetic field, and a properly chosen field
can resonantly couple colliding atoms into the molecular state.
The inherent difference in magnetic moments facilitates a
Stern-Gerlach selection of molecules and atoms. Two recent
experiments \cite{Herbig2003,Duerr2003} demonstrate the separation
of the molecular from the atomic cloud in free space.

In this Letter, we report the creation of a pure sample of up to
$3\times10^5$ optically trapped molecules from a fermionic gas of
$^6$Li atoms. After the production of an atom-molecule mixture via
three-body collisions, a Stern-Gerlach purification scheme
efficiently removes all trapped atoms, while leaving all molecules
trapped. This allows us to investigate the intriguing behavior of
the pure molecular sample, which strongly depends on the magnetic
field.

\begin{figure}
\includegraphics[width=8cm]{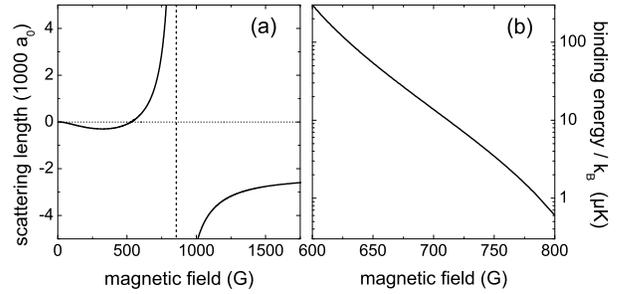}
\vspace{-2mm} \caption{\label{resonance} (a) Magnetic-field
dependence of the s-wave scattering length $a$ in the $^6$Li spin
mixture. An additional, narrow Feshbach resonance at $\sim$550\,G
\cite{O'Hara2002b} is omitted in the plot. (b) Binding energy of
the weakly bound molecular level in the region of large positive
$a$.}
\end{figure}

The lithium isotope $^6$Li is one of the two prime candidates in
current experiments exploring the physics of fermionic quantum
gases
\cite{Truscott2001a,O'Hara2002a,Bourdel2003a,Gupta2003a,Jochim2002a},
the other one being $^{40}$K \cite{Regal2003a,Modugno2002a}. A
spin mixture composed of the lowest two sublevels in the hyperfine
manifold of the electronic ground state \cite{note_twostates} is
stable against two-body decay and exhibits wide magnetic
tunability of s-wave interactions via a broad Feshbach resonance
at about 850\,G \cite{Houbiers1998a}. A calculation of the
corresponding scattering length $a$ as a function of the magnetic
field \cite{O'Hara2002b} is shown in Fig.~\ref{resonance}(a)
\cite{resonance_note}. The large cross section for elastic
scattering near the resonance can be used for efficient
evaporative cooling, in particular above the resonance at negative
scattering length where inelastic loss is neglible
\cite{O'Hara2002a}. In the region of positive scattering length
below the resonance, loss features have been observed
\cite{Dieckmann2002a}. At large positive $a$, a weakly bound
molecular level exists with a binding energy approximately given
by $\hbar/(ma^2)$, where $\hbar$ is Planck's constant and $m$
denotes the atomic mass. For the region of interest,
Fig.~\ref{resonance}(b) shows this binding energy as calculated
from the scattering length data \cite{bindingenergy}.


The starting point of our experiments is a sample of $2.5 \times
10^6$ $^6$Li atoms in a standing-wave optical dipole trap realized
with a Nd:YAG laser at a wavelength of 1064\,nm
\cite{Mosk2001,Jochim2002a}. The 50-50 spin mixture in the lowest
two spin states \cite{note_twostates} is spread over $\sim$1500
individual lattice sites of the standing wave trap. In the central
region of the trap, a single site contains typically 1800 atoms.
The axial and radial trap frequencies are 390\,Hz and 260\,kHz,
respectively. The trap depth is $k_{\rm B} \times 27\,\mu$K with
$k_{\rm B}$ denoting Boltzmann's constant. At a temperature of
2.5\,$\mu$K peak values for the number density and phase-space
density are $3 \times 10^{12}$\,cm$^{-3}$ and 0.04
\cite{errornote,2Dnote}, respectively. The ultracold gas is
prepared by forced evaporative cooling after loading the optical
trap at an initial depth of $\sim$1\,mK with $8 \times 10^6$ atoms
from a magneto-optical trap (MOT). The evaporation is performed by
ramping down the light intensity in 1\,s at a magnetic field of
1200\,G. The evaporation initially proceeds with very high
efficiency similarly to \cite{granade2002,O'Hara2002a}, but
finally loses its efficiency when the tigthly confining lattice
potential does not support more than one or two quantum states
\cite{note_tunneling}.

We form molecules at a field of 690\,G, where we find optimum
production rates at a large positive scattering length of
$a=+1300a_0$. Here $a_0$ denotes Bohr's radius. To reach the
production field of 690\,G we quickly ramp from the evaporation
field of 1200\,G down to this value with a speed of $-7.5$\,G/ms.
In contrast to other experiments with fermionic atoms
\cite{Regal2003a,Cubizolles2003}, the molecule formation during
this ramp is negligible and the molecules are predominantly formed
after the ramp at the fixed production field.

The molecules are detected by dissociating them into atoms
\cite{Regal2003a,Herbig2003,Duerr2003,Cubizolles2003} and
measuring their fluorescence. For this purpose, we apply a ramp
across the Feshbach resonance to fields of typically 1200\,G
(speed $+6\,$G/ms). This brings the bound level above the
scattering continuum and the molecules quickly dissociate. The
dissociation turns out to be insensitive to variations of the ramp
speed and the final field. After the dissociation ramp, we
immediately ramp down to zero magnetic field. The ramp speed of
$-12$\,G/ms is fast enough to avoid molecule formation when
crossing the region of positive scattering length. After reaching
zero magnetic field we recapture all atoms into the MOT. Their
number is then determined by measuring the emitted fluorescence
intensity using a calibrated photodiode \cite{errornote}. This
measurement provides the total atom number $2N_{\rm mol}+N_{\rm
at}$, where $N_{\rm mol}$ and $N_{\rm at}$ denote the number of
molecules and atoms after the production phase, respectively. To
determine $N_{\rm at}$ we repeat the same measurement without the
Feshbach dissociation ramp by immediately ramping down to zero
from the production field. The ramp down to zero magnetic field
increases the binding energy to a large value of about $k_{\rm
B}\times$80\,mK and the molecules are lost without leading to any
fluorescence light in the MOT. The number of molecules $N_{\rm
mol}$ is then obtained by taking the difference in atom numbers
measured in two subsequent runs with and without the dissociating
Feshbach ramp.

\begin{figure}
\includegraphics[width=8cm]{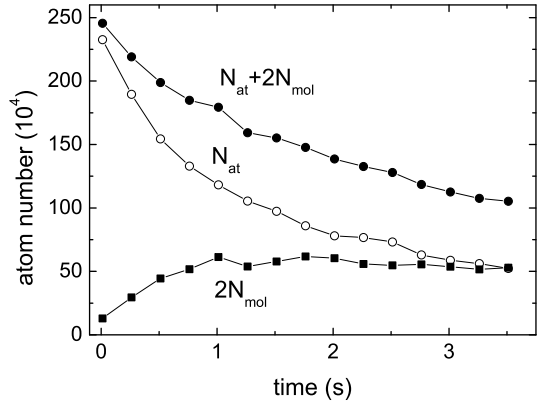}
\vspace{-2mm}
\caption{\label{creation} Formation of molecules at a
fixed magnetic field of 690\,G. The measured numbers $N_{\rm at} +
2N_{\rm mol}$ and $N_{\rm at}$ are plotted as a function of time
together with the resulting number of molecules $2N_{\rm mol}$.}
\end{figure}

The creation of molecules from the atomic gas is demonstrated in
Fig.~\ref{creation} for the optimum production field of 690\,G.
The time evolution of the measured numbers $2N_{\rm mol} + N_{\rm
at}$ and $N_{\rm at}$ is shown together with the corresponding
number of molecules $2N_{\rm mol}$. We attribute the molecule
formation to three-body recombination
\cite{Suno2003a,Petrov2003a}. Two-body processes cannot lead to
bound dimers as a third particle is required for energy and
momentum conservation. The three-body molecule formation process
can be modeled with the differential equation $\dot{N}_{\rm
mol}/N_{\rm at} = M_3 \langle n_{\rm at}^2\rangle$, where $\langle
n_{\rm at}^2\rangle$ denotes the mean quadratic density of the
atoms. From the initial molecule formation rate of $\dot{N}_{\rm
mol} = 3.5 \times 10^5$\,s$^{-1}$ we thus derive a three-body
formation coefficient of $M_3 = 1 \times 10^{-25}$cm$^6$/s$^{-1}$
\cite{errornote}.
The maximum number of $3 \times 10^5$ molecules is reached after
about 1\,s. For longer times, the fraction of atoms forming
molecules approaches a value of $\sim$50\%.

At the optimum production field of 690\,G the molecular binding
energy amounts to $\sim$$k_{\rm B}\times 18\,\mu$K, which is in
between the thermal energy of $k_{\rm B}\times2.5\,\mu$K and the
trap depth of $k_{\rm B}\times 27\,\mu$K for the atoms. For the
molecules the trap depth is a factor of two higher because of the
two times larger polarizability. We have verified this fact by
measuring the trap frequencies for atoms and molecules to be equal
within the experimental uncertainty of a few percent. After a
three-body recombination event both the atom and the molecule
remain trapped. We believe that the recombination heat is cooled
away by a evaporation of atoms out of the trap. Evaporative loss
of molecules is strongly suppressed because of the higher trap
depth.

\begin{figure}
\includegraphics[width=8cm]{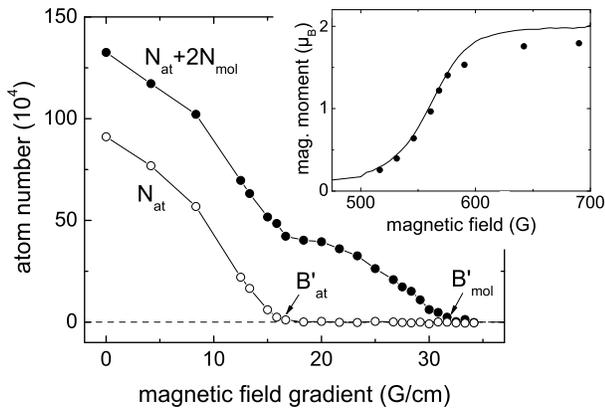}
\vspace{-2mm}
\caption{\label{sterngerlach} Stern-Gerlach selection
by applying a magnetic field gradient to the trapped atom-molecule
mixture at 568\,G and a trap depth of $k_{\rm B} \times 19\,\mu$K.
Marked are the two gradients where all the atoms and all the
molecules are lost. The inset shows the magnetic moment of the
molecules estimated from the Stern-Gerlach selection at different
magnetic fields together with the theoretical calculation.}
\end{figure}

To purify the created molecules we use a Stern-Gerlach selection
technique. We apply a magnetic field gradient perpendicular to the
standing wave axis. This pulls particles out of the trap for which
the magnetic force is larger than the trapping force. In order to
be able to apply large enough field gradients, we lower the trap
depth to k$_{\rm B}\times$19\,$\mu$K while applying the gradient
for about 10\,ms. Fig.~\ref{sterngerlach} demonstrates such a
purification at 568\,G. While all the atoms are lost above
$B^{\prime}_{\rm at}=17$\,G/cm, the molecules start getting
spilled at 20\,G/cm, and are lost completely above
$B^{\prime}_{\rm mol}=32.5$\,G/cm. This means that under suitable
conditions, we can remove all the atoms while keeping the molecule
number constant.

The magnetic moment of the molecules $\mu_{\rm mol}$ can be
estimated to be \mbox{$\mu_{\rm mol}=2 \mu_{\rm at}
B^{\prime}_{\rm mol}/ B^{\prime}_{\rm at}$}, where $\mu_{\rm at}$
is the magnetic moment of one free atom. At high magnetic field,
$\mu_{\rm at}$ equals Bohr's magneton $\mu_{\rm B}$. The inset of
Fig.~\ref{sterngerlach} shows the magnetic moments of the
molecules determined at various magnetic fields. The data agree
well with the magnetic field dependence calculated from theory
(solid curve). We attribute the systematic deviation to slightly
different trap parameters for atoms and molecules.

\begin{figure}
\includegraphics[width=8cm]{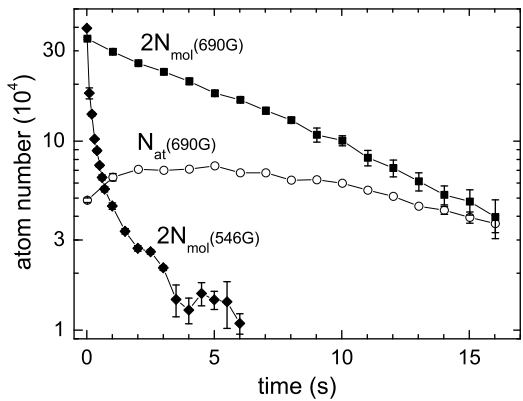}
\vspace{-2mm}
\caption{\label{lifetime} Time evolution of an
initially pure sample of molecules at 546\,G
({\scriptsize$\blacklozenge$}) and at 690\,G
({\tiny$\blacksquare$}). At 690\,G atoms are observed to reappear
({\large$\circ$}).}
\end{figure}

Starting with a pure molecular sample, we study its stability
against inelastic molecule-molecule collisions. Corresponding
decay curves are displayed in Fig.~\ref{lifetime} for two
different magnetic fields. At 546\,G a rapid non-exponential decay
is observed as a clear signature of inelastic molecule-molecule
collisions. From the initial decay rate we derive a two-body loss
coefficient of $5 \times 10^{-11}$\,cm$^3$/s \cite{errornote}. At
690\,G, the observed behavior is strikingly different. The
molecular sample shows a nearly exponential decay with a time
constant as long as $\sim$10\,s. As similar lifetimes are observed
for trapped atom samples under conditions where trapped molecules
cannot be created, the observed molecular lifetime can be fully
attributed to one-body effects like heating in the optical trap.
For a loss rate coefficient at 690\,G our data provide an upper
limit of $3 \times 10^{-13}$\,cm$^3$/s \cite{errornote}, which is
surprisingly low for inelastic collisions in a molecular system
with many open exit channels.

The data at 690\,G show another interesting collisional effect.
Atoms reappear after purification of the molecular cloud, see
({\large$\circ$}) in Fig.~\ref{lifetime}. For long storage times
($\sim$15\,s) an atom-molecule mixture is reestablished with a
similar fraction of molecules as observed in the initial formation
process at the same magnetic field, see Fig.~\ref{creation}.
Collisions producing atoms from molecules are endoergic in nature
as kinetic energy is required to provide the dissociation energy.
The increasing atom fraction does not lead to any increased loss.
This shows that the gas is remarkably stable both against
molecule-molecule and atom-molecule collisions.

\begin{figure}
\includegraphics[width=8cm]{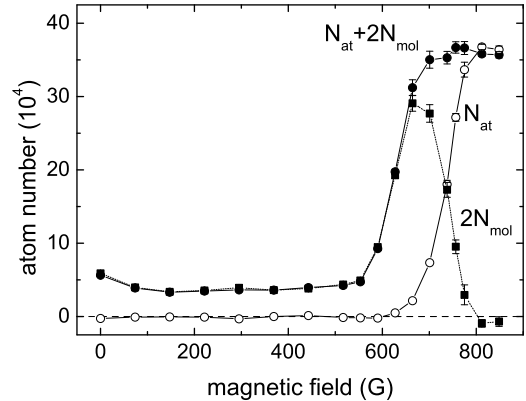}
\vspace{-2mm}
\caption{\label{magfieldsurvey} Remaining number of
atoms $N_{\rm at}$, $N_{\rm at} + 2 N_{\rm mol}$ and $2 N_{\rm
mol}$ after a 1-s hold time at variable magnetic field starting
with a pure molecular sample.}
\end{figure}

The dependence of the molecular decay on the magnetic field is
shown in Fig.~\ref{magfieldsurvey}. Here we store the initially
pure gas of $1.8 \times 10^5$ molecules at a variable magnetic
field for a fixed holding time of 1\,s before we measure the
number of remaining molecules and atoms. A sharp transition occurs
around 650\,G. For fields below $\sim$600\,G, where the binding
energy is relatively large ($> k_{\rm B} \times 100\,\mu$K), the
observed decay is very fast and no atoms are found to reappear.
Here inelastic collisions apparently lead to a rapid vibrational
quenching. Furthermore, the kinetic energy of the molecules cannot
provide the necessary energy for collisional dissociation.
Consequently, we do not observe any atoms reappearing.

For fields above $\sim$680\,G a completely different behavior is
observed. In this regime, no significant loss occurs in the total
number $2N_{\rm mol}+N_{\rm at}$. However, an increasing atom
fraction is observed as a result of collisional dissociation of
the molecules. Here the binding energy approaches the thermal
energy and the sample tends towards a thermal atom-molecule
equilibrium. Close to the Feshbach resonance, where the binding
energy becomes comparable to thermal energy, the atomic fraction
dominates in the atom-molecule mixture.

In conclusion we have produced an ultracold, pure molecular gas of
$^6$Li dimers in an optical dipole trap. Close to the Feshbach
resonance where the molecular binding energy is small there is a
strong coupling of the atomic gas and the molecules. Three-body
collisions between atoms form molecules and collisions break up
molecules to produce atoms. Our observations show that this
exchange between atomic and molecular fraction can be nearly
lossless. The long molecular lifetime along with a large elastic
collision rate between the particles opens up great perspectives
for further evaporative cooling of the molecular gas to
Bose-Einstein condensation. Given the maximum molecule number of
3$\times$10$^5$ and a temperature of about 2.5\,$\mu$K we reach a
phase space density of 0.01, only a factor of 4 lower than our
initial atomic phase space density. The molecular sample may be
further cooled to condensation by efficient evaporation. Out of a
mixture of atoms and molecules, mainly atoms will evaporate
because they are more weakly trapped than the molecules.  The gas
is cooled further when molecules break up into atoms since this is
an endoergic process. Once quantum degeneracy is accomplished it
will be very interesting to cross the Feshbach resonance in order
to observe the transition to a strongly interacting superfluid
Fermi gas
\cite{Holland2001a,Timmermans2001a,Chiofalo2002,Ohashi2002}.

\begin{acknowledgments}
We thank G.\ Shlyapnikov for very stimulating discussions and V.\
Venturi for providing us with theoretical data on the scattering
length and binding energy. We gratefully acknowledge support by
the Austrian Science Fund (FWF) within SFB 15 (project part 15)
and by the European Union in the frame of the Cold Molecules TMR
Network under contract No. HPRN-CT-2002-00290.
\end{acknowledgments}



\begin{thebibliography}{33}
\bibitem[{not({\natexlab{a}})}]{note_twostates}
\bibinfo{note}{The lowest two atomic states of the $^2$S$_{1/2}$ ground state
  are characterized by the quantum numbers $F=1/2,m_F=1/2$ $(m_F=-1/2)$ at low
  magnetic fields, and $m_J=-1/2, m_I=+1$ $(m_I=0)$ at high fields.}
\bibitem[{err()}]{errornote}
\bibinfo{note}{With fluorescence detection we can determine the absolute atom
  number only within a factor of two. All the derived quantities are subject to
  this systematic error.}
\bibitem[{2Dn()}]{2Dnote}
\bibinfo{note}{With a ground-state population of 99\% in the tightly confined
  direction each lattice site contains a quasi-2D gas. This fact is taken into
  account in our calculations of number and phase-space densities.}
\bibitem[{not({\natexlab{b}})}]{note_tunneling}
\bibinfo{note}{We believe that the final limitation of evaporative cooling in
  our trap is due to tunneling loss.}
\end{thebibliography}

\end{document}